\documentclass[a4paper,11pt]{article}
\usepackage{pos}
\usepackage{subfigure} 
\usepackage[absolute,overlay]{textpos} 

\graphicspath{{./figs/}}

\newcommand{\qslash}[1]{\text{$\not \! #1$}}

\newcommand{\CO}{\mathcal{O}}
\newcommand{\GeV}{\mathop{\rm GeV}\nolimits}

\newcommand{\MSb}{{\overline{\text{MS}}}}
\DeclareMathOperator{\Tr}{Tr}

\begin{document}

\title{Update on flavor diagonal nucleon charges from clover fermions}

\begin{textblock}{20}(14.0,1.70)
LLNL-PROC-858664
\end{textblock}

\author*[a,b,c]{Sungwoo Park}
\author[d]{Tanmoy Bhattacharya}
\author[d]{Rajan Gupta}
\author[f]{Huey-Wen Lin}
\author[e]{Santanu Mondal}
\author[g]{Boram Yoon}
\affiliation[a]{Physical and Life Sciences Division, Lawrence Livermore National Laboratory, Livermore, CA 94550, USA}
\affiliation[b]{Nuclear Science Division, Lawrence Berkeley National Laboratory, Berkeley, CA 94720, USA}
\affiliation[c]{Thomas Jefferson National Accelerator Facility,
  12000 Jefferson Avenue, Newport News, VA 23606, USA}
\affiliation[d]{Theoretical Division T-2, Los Alamos National Laboratory, Los Alamos, NM 87545, USA}
\affiliation[e]{Ibsyn Scientific, 75C Park St, Kolkata, India 700016}
\affiliation[f]{Department of Physics and Astronomy, Michigan State University, MI, 48824, USA}
\affiliation[g]{NVIDIA Corporation, Santa Clara, CA 95051, USA}

\emailAdd{park49@llnl.gov}
\emailAdd{tanmoy@lanl.gov}
\emailAdd{rajan@lanl.gov}
\emailAdd{hwlin@pa.msu.edu}
\emailAdd{santanu.sinp@gmail.com}
\emailAdd{byoon@nvidia.com}

\abstract{We present a summary of the full calculation of the axial, scalar and tensor flavor diagonal charges of the nucleon carried out using Wilson-clover fermions on eight ensembles generated using 2+1+1-flavors of highly improved staggered quarks (HISQ) by the MILC collaboration.  We also give results for the $3\times 3$ matrix of renormalization factors between the RI-sMOM and $\overline{\rm MS}$ scheme for the 2+1 flavor theory that include flavor mixing. Preliminary results for $g_{A,S,T}^{u,d,s}$ are presented in the $\overline{\rm MS}$ scheme at scale 2~GeV. }

\FullConference{The 40th International Symposium on Lattice Field Theory (Lattice 2023)\\
July 31st - August 4th, 2023\\
Fermi National Accelerator Laboratory\\}

\maketitle

\section{Introduction}

We present lattice QCD results for flavor diagonal nucleon charges
$g_{A,S,T}^{u,d,s}$ extracted from the matrix elements, within 
ground state nucleons, of axial,
scalar, and tensor quark bilinear operators, ${\overline q}\Gamma q$
with the Dirac matrix $\Gamma = \gamma_\mu \gamma_5, I,
\sigma_{\mu\nu}$, respectively.  The
calculations were done using Wilson-clover fermions on eight ensembles
generated using 2+1+1-flavors of highly improved staggered quarks
(HISQ) by the MILC collaboration \cite{Bazavov:2012xda}.
The motivation for these calculations and much of the methodology used
has already been published for $g_{A}^{q}$ in Ref.~\cite{Lin:2018obj},
$g_T^{q}$ in \cite{Gupta:2018lvp} and the the pion-nucleon sigma term,
$\sigma_{\pi N}=m_{u,d}\times g_S^{u+d}$, in~\cite{Gupta:2021ahb}.  A
review of these quantities calculated until 2021 by various lattice
collaborations has been presented in the latest FLAG report
2021~\cite{FLAG:2021npn}. Here, we focus on the progress since Lattice
2022~\cite{Park:2023tsj}, in particular the full  
nonperturbative determination of renormalization factors including 
flavor mixing used to get $g_{A,S,T}^{u,d,s}$ in the $\overline{\rm MS}$ scheme at scale 2~GeV.\looseness-1

\section{Nonperturbative renormalization}
\label{sec:NPR}

We calculate the renormalization constants for the flavor diagonal
bilinear operators $\CO^f= \bar \psi^f \Gamma \psi^f$ with Dirac
matrix $\Gamma$ and the flavor index $f=\{u,d,s\}$ in the $N_f=3$
theory. The general relation between renormalized, $\CO_R$, and bare,
$\CO$, operators including mixing between flavors is given by $\CO^f_R
= \sum_{f'} Z_\Gamma ^{ff'}\CO^{f'}$.  We determine $Z_\Gamma^{ff'}$
nonperturbatively on the lattice using the regularization independent
(RI) renormalization scheme~\cite{Martinelli:1994ty} in which the
renormalized vertex function is set to its tree-level value.  The
calculation is done with the gauge fields fixed to the landau gauge.

\subsection{Flavor mixing in the RI scheme}
\begin{figure}[t]
  \center
  \subfigure[The connected vertex $\Gamma_\text{conn}(m,\mu^2)$]{ \includegraphics[width=0.45\textwidth]{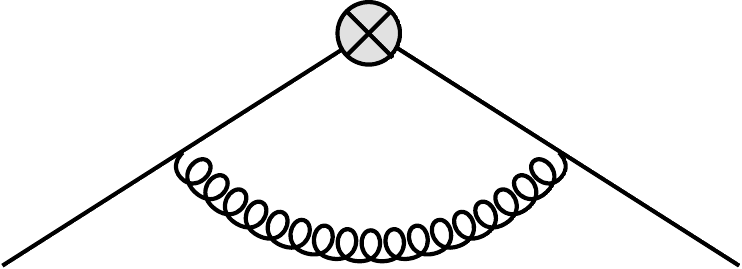}}\hspace{0.3in}
  \subfigure[The disconnected vertex $\Gamma_\text{disc}(m_f,m_{f'},\mu^2)$]{ \includegraphics[width=0.45\textwidth]{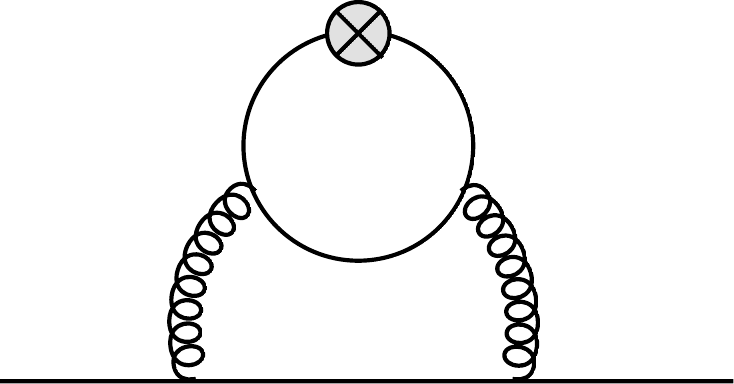}}
  \caption{The amputated vertex diagrams defined in Eq.~\protect\eqref{eq:AVF}}
  \label{fig:diagram}
\end{figure}

We start with the amputated vertex function $\Gamma^{ff'}(p_1,p_2)$
defined as,
\begin{align}
  \Gamma^{ff'}(p_1,p_2) &= \langle
  S^f(p_1)\rangle^{-1} \langle \psi^f(p_1) \CO^{f'} \psi^f(p_2) \rangle
  \langle S^f(p_2)\rangle^{-1}
  \label{eq:AVF}
\end{align}
where $\psi^f$ and $S^f$ are the quark field and the propagator
with flavor-$f$, and each $\langle \cdots \rangle$ is color traced. 
The three-point function $\langle \psi^f(p_1) \CO^{f'}
\psi^f(p_2) \rangle$ has both connected and disconnected
diagrams shown in Fig.~\ref{fig:diagram}. With the wave function renormalization $Z_\psi^f$ 
defined by $(Z_\psi^f)^{1/2} \psi^f=\psi_R^f$, the renormalized
amputated vertex function is 
\begin{align}
  \Gamma^{ff'}_R(p_1,p_2)
  &=\frac{Z_\Gamma^{f'f''}}{Z_\psi^f}\Gamma^{ff''}(p_1,p_2) \,,
  \label{eq:Gamma}
\end{align}
which defines the renormalization (including mixing) matrix $Z_\Gamma$.
Next we do a spin trace using a projection operator $\mathbb{P}$ chosen appropriately depending on the Dirac structure and momentum of $O$ to give the projected amputated vertex function
$\Lambda^{ff'}\equiv \text{Tr}\left[\mathbb{P}\ 
  \Gamma^{ff'}\right]$, and the RI condition $\Lambda^{ff'}_R \equiv \text{Tr}\left[\mathbb{P}\ 
  \Gamma_R^{ff'}\right] =\delta^{ff'}$
fixes it to its tree-level value. Lastly we absorb $Z_\psi$ by defining 
$\widetilde{\Lambda}^{ff''}\equiv (Z_\psi^{f})^{-1}\Lambda^{ff''}$ and get a simple form for the $N_f \times N_f$ flavor renormalization matrix,
\begin{align}
  Z_\Gamma^{ff'}=\left[(\widetilde{\Lambda}^T)^{-1} \right]^{ff'}.
\end{align}
The determination of $Z_\psi^{f}$ connecting $\Lambda$ and $\widetilde{\Lambda}$ is done using two methods discussed later.


The amputated vertex function $\Gamma^{ff'} = \Gamma_\text{conn}(m^f,\mu^2)\delta^{ff'} -
  \Gamma_\text{disc}(m^f,m^{f'},\mu^2)$ gets contributions from both the connected
and disconnected diagrams shown in Fig.~\ref{fig:diagram}. 
The (-) sign in $\Gamma_\text{disc}$ is due to the
anticommuting nature of the fermion fields, i.e., it accounts for the quark loop. The projected amputated vertex functions $c_f$ and $d_{ff'}$, including the factor $(Z_\psi^f)^{-1}$, are given by
\begin{align}
  \widetilde{c}_\Gamma^f &= \frac{1}{Z_\psi^f}\text{Tr}\left[\mathbb{P}_{\cal C} \Gamma_\text{conn}(m^f,\mu^2) \right] = \frac{{c}_\Gamma^f}{{Z_\psi^f}}
  & \widetilde{d}_\Gamma^{ff'} &=
  \frac{1}{Z_\psi^f}\text{Tr}\left[\mathbb{P}_{\cal C} \Gamma_\text{disc}(m_f,m_{f'},\mu^2) \right] = \frac{{d}_\Gamma^{ff'}}{{Z_\psi^f}}.
    \label{eq:cd-def}
\end{align}

For the $N_f=2+1$ isospin symmetric theory relevant to this work, the determination of 
$Z^{ff'}$ requires calculating the following 6 quantities,
\begin{align}
  c_\Gamma^l,~c_\Gamma^s,~d_\Gamma^{ll},~d_\Gamma^{ls},~d_\Gamma^{sl}\text{, and }d_\Gamma^{ss}.
\end{align}
Working in the flavor basis $f\in \{u+d,u-d,s\}$, $ Z_\Gamma$
becomes block diagonal:
\begin{align}
    Z_\Gamma&=
  \begin{pmatrix}
    Z_\Gamma^{u-d,u-d} & 0 & 0 \\
    0 & Z_\Gamma^{u+d,u+d} & Z_\Gamma^{u+d,s}\\
    0 & Z_\Gamma^{s,u+d} & Z_\Gamma^{ss} \\
  \end{pmatrix}
  =
  \begin{pmatrix}
    \widetilde{c}_\Gamma^l & 0 & 0 \\
    0 & \widetilde{c}_\Gamma^l-2\widetilde{d}_\Gamma^{ll} & -2\widetilde{d}_\Gamma^{sl} \\
    0 & -\widetilde{d}_\Gamma^{ls} & \widetilde{c}_\Gamma^s-\widetilde{d}_\Gamma^{ss}\\
  \end{pmatrix}^{-1}.
  \label{eq:Z_RI}
\end{align}

\subsection{RI-sMOM scheme}
\begin{figure*}[h]
  \center
  \includegraphics[width=0.3\linewidth]{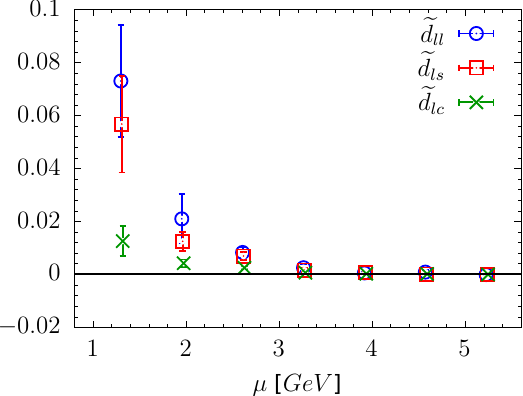}
  \includegraphics[width=0.3\linewidth]{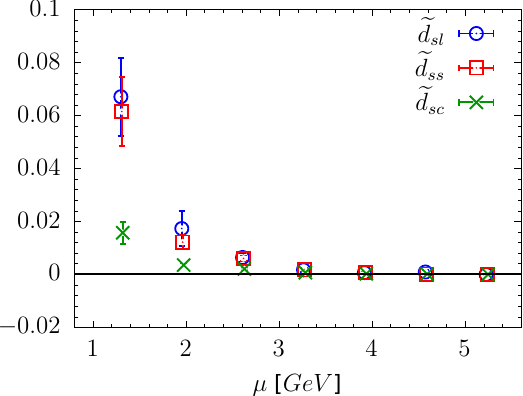}
  \includegraphics[width=0.3\linewidth]{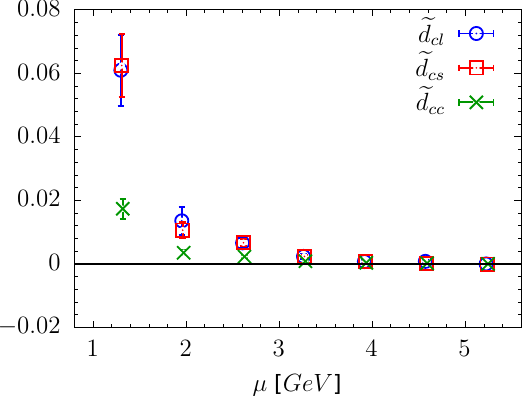}
  \caption{Disconnected projected amputated Green's function
    $\widetilde{d}_\Gamma^{ff'}$ with quark flavors $f$ and $f'$ for a scalar
    bilinear operator $O_S^{f'}$ at $M_\pi=220$MeV ensemble. Here,
    $\mu^2=p^2={p'}^2=(p-p')^2$ in the RI-sMOM scheme.
  }
  \label{fig:d_lf}
\end{figure*}

Calculation of $Z_\Gamma$ is done in the RI-sMOM 
scheme \cite{Sturm:2009kb} in which the
4-momentum of the external legs $\{p_1,p_2\}$ satisfies the symmetric
momentum condition $p_1^2=p_2^2=(p_1-p_2)^2=\mu^2$ with $\mu^2$
defining the renormalization scale. We find that the 
matrix $Z^\text{RI-sMOM}(\mu)$ is close to diagonal with $c_\Gamma^l\sim
O(1)$ and $d_\Gamma^{ff'}$ at most a few percent at $\mu\gtrsim2$~GeV.
This is illustrated in Fig.~\ref{fig:d_lf} for the scalar charge
(largest mixing) for 3 disconnected projected amputated Green's
function $d_\Gamma^{ff'}$, including $f' = {\rm charm}$, calculated at various $\mu$. The
value decreases as the quark mass in the loop is increased from light
to strange to charm, becoming subpercent for charm for $\mu \gtrsim 2\GeV$.
To get a signal for such small mixing, we use the momentum
source method and choose the momenta $\{p_1,p_2\}$  to minimize $O(4)$-symmetry breaking. An example is $p_1 = (1,1,1,1)$ and $p_2 = (1,1,1,-1)$.

\subsection{Matching and RG running}

At each value of the scale $\mu^2=p_i^2=q^2$, the renormalization factors
(mixing matrix elements) in RI-sMOM scheme are matched perturbatively~\cite{Gracey:2011fb} to $\MSb$ scheme, $C^{\text{RI-sMOM}\to \MSb}(\mu) = \frac{Z^\text{RI}
  (\mu)}{Z^\MSb (\mu)}$.  
One of the advantages of using RI-sMOM
scheme for the scalar channel is the better convergence in the
perturbative series of the conversion factor compared to the RI-MOM
scheme \cite{Chetyrkin:1999pq, Sturm:2009kb}.

Flavor nonsinglet axial current is conserved and therefore there is no
matching from RI to $\MSb$ scheme nor RG running due to the Ward
identity. On the other hand, conservation of the flavor singlet current is broken by the chiral anomaly, and the renormalization constant
becomes nontrivial at the two-loop level. Here, we use the 3-loop anomalous
dimension \cite{Larin:1993tq} for the RG running. On the
other hand, the matching to $\MSb$ is $1+\CO(\alpha^2)$
\cite{Green:2017keo, Gracey:2020rok} and we drop the two-loop contribution that is expected to be subpercent.
For scalar and tensor operators, two-loop conversion to $\MSb$
\cite{Gracey:2011fb} and three-loop running 
\cite{Chetyrkin:1999pq, Gracey:2000am} is used.

Remaining  dependence on the RI-sMOM scale is 
removed by fits using the quadratic ansatz, \\
$Z^{\MSb}(2\GeV;\mu)=Z^{\MSb}(2\GeV) + c_1 \mu^2 + c_2 \mu^4$. 
We do not have enough 
data points at small $\mu^2$ where non-perturbative effects can be large. So we make fits with a large lower value of
$\mu$, and no longer include a $1/\mu^2$ term in the ansatz
as was done in Refs.~\cite{Bhattacharya:2016zcn,Yoon:2016jzj}.

\subsection{Renormalization Strategies \texorpdfstring{$\rm{Z_1}$}{Z1} and \texorpdfstring{$\rm{Z_2}$}{Z2} based on  \texorpdfstring{$Z_\psi^q$}{Zpsi}}
\label{sec:Z1Z2}

The renormalization constants for the \emph{isovector} bilinear
operators in the RI scheme is given by $Z_\Gamma|_{\rm
  RI}(p)=\frac{Z_\psi(p)}{c_\Gamma(p)}$, where $\sqrt{Z_\psi}$ is the
renormalization constant for the fermion field and $c_\Gamma$ is the
projected amputated connected 3-point function calculated in Landau
gauge and defined in Eq.~\eqref{eq:cd-def}.

We calculate $Z_\psi$ in two ways, which define the two 
renormalization strategies $Z_1$ and $Z_2$:
\begin{itemize}
  
\item 
$\rm Z_1$: $Z_\psi$ is calculated from the projected bare quark
  propagator, $Z_\psi(p)=\frac{i}{12p^2}\Tr[S_B^{-1}(p)\qslash{p}]$
\item $\rm Z_2$: We use $Z_\psi^{\rm VWI}(p) = c_V(p)/g_V$, where $c_V$
  is the projected amputated connected 3-point function with insertion
  of the local vector operator within the quark state while the isovector
  vector charge   $g_V$ is from insertion of the vector current within 
  the nucleon state. Using the vector 
  Ward identity (VWI)  $Z_V g_V=1$ implies $Z_\Gamma|_{\rm Z_2} = c_V/c_\Gamma$.
\end{itemize}

\begin{figure}[t]
  \center
  \includegraphics[width=0.48\textwidth]{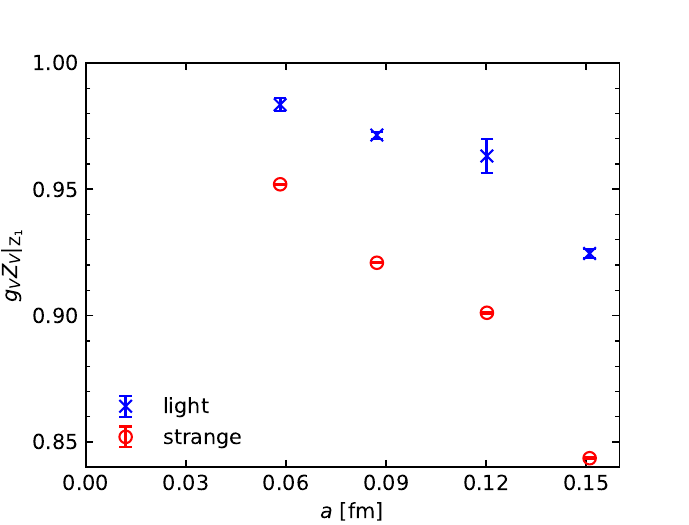}
  \caption{$g_V^f Z_V^f|_{\rm Z_1}$ for the light and strange valence quark
    masses. The isovector vector charge $g_V^f$ is from the forward matrix
    element with a nucleon state for both quark masses. The vector renormalization factor is
    $Z^f_V|_{\rm Z_1}=Z_\psi^f/c^f_V$ for a flavor $f\in\{l,s\}$. The vector
    Ward Identity (VWI), $g_V Z_V|_{\rm Z_1}=1$ is likely restored in the
    continuum limit, however, a linear extrapolation is insufficient.}
  \label{fig:gVZV}
\end{figure}

These two strategies were used in Ref.~\cite{Bhattacharya:2016zcn,Yoon:2016jzj} 
for \emph{isovector} bilinear operators of light quarks, i.e., 
$g_\Gamma Z_\Gamma$ and $Z_\Gamma/Z_V \times g_\Gamma/g_V$. Later, they were 
called $\rm Z_1$ and $\rm Z_2$  in Ref.~\cite{Park:2021ypf}. 
In Ref.~\cite{Bhattacharya:2016zcn,Yoon:2016jzj} we showed that they have
different behavior versus $q$ as the various discretization effects are
different in $Z_\psi$ and $Z_\psi^{\rm VWI}$ obtained from $S_B(p)$
and $c_V(p)/g_V$, respectively.
Our data show \looseness-1
\begin{itemize}
\item For both the light and strange quarks, $Z_V|_{\rm Z_1}\times g_V^{l,{\rm bare}} \to 1$ as $a \to 0$, however the deviation  at a given $a$ increases for $m_l \to m_s$ (a significant mass effect due to discretization) as shown
  in Fig.~\ref{fig:gVZV}.  Satisfying this relation in the continuum limit implies that renormalization using $\rm  Z_1$ and $Z_2$ will give consistent results.
\item Final continuum limit values of nucleon charges and form factors
  using $\rm Z_1$ and $\rm Z_2$ agree within the quoted
  errors~\cite{Park:2021ypf}.
\end{itemize}

\begin{figure*}
  \center
  \includegraphics[width=0.24\textwidth]{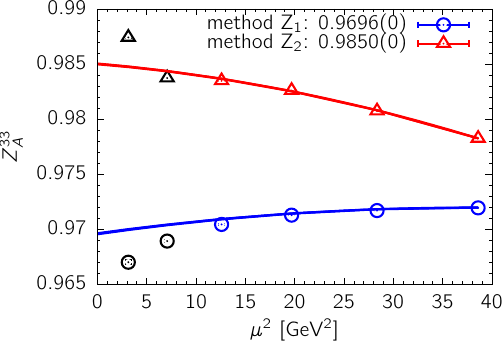}
  \includegraphics[width=0.24\textwidth]{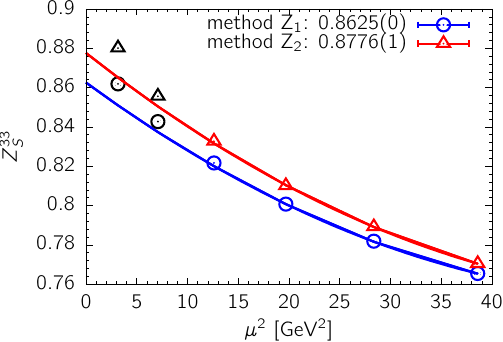}
  \includegraphics[width=0.24\textwidth]{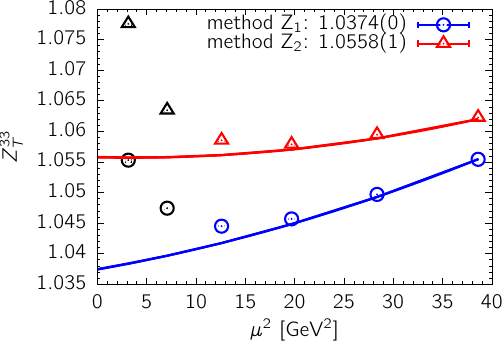}
  \includegraphics[width=0.24\textwidth]{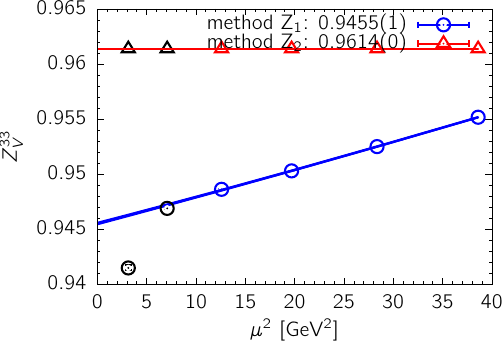}

  \includegraphics[width=0.24\textwidth]{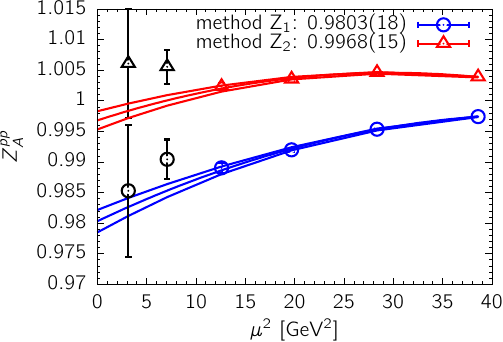}
  \includegraphics[width=0.24\textwidth]{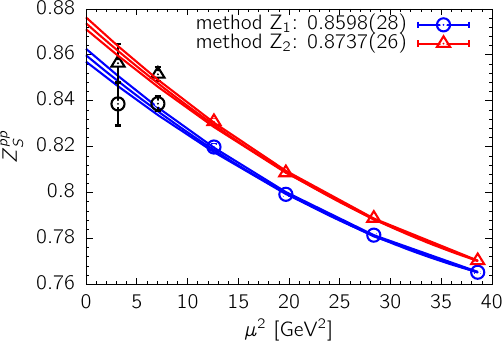}
  \includegraphics[width=0.24\textwidth]{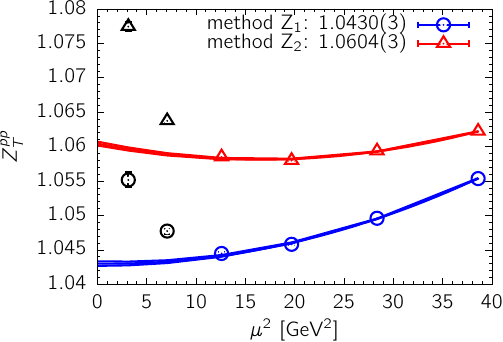}
  \includegraphics[width=0.24\textwidth]{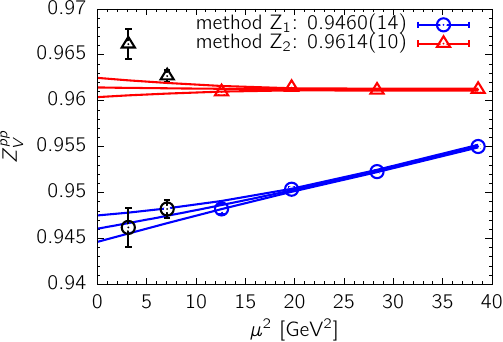}

  \includegraphics[width=0.24\textwidth]{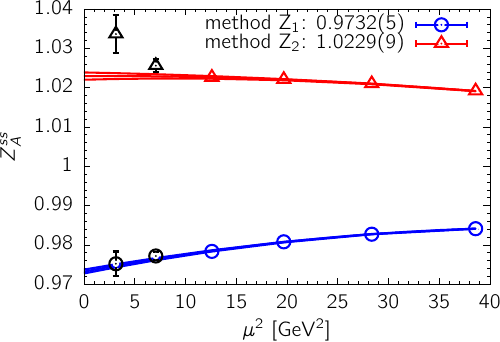}
  \includegraphics[width=0.24\textwidth]{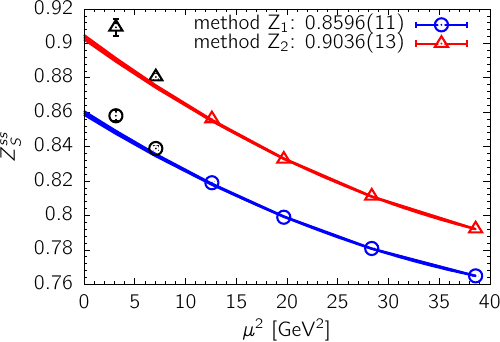}
  \includegraphics[width=0.24\textwidth]{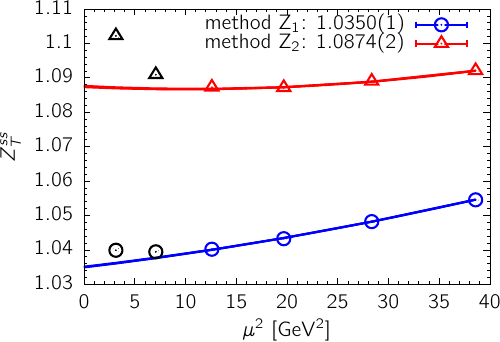}
  \includegraphics[width=0.24\textwidth]{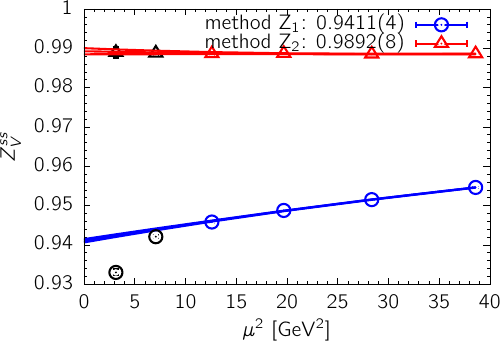}

  \includegraphics[width=0.24\textwidth]{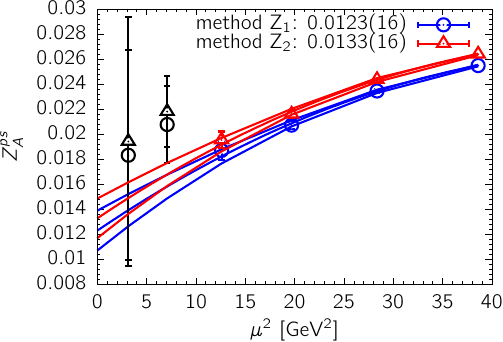}
  \includegraphics[width=0.24\textwidth]{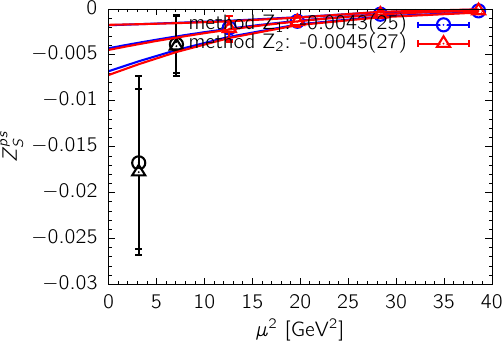}
  \includegraphics[width=0.24\textwidth]{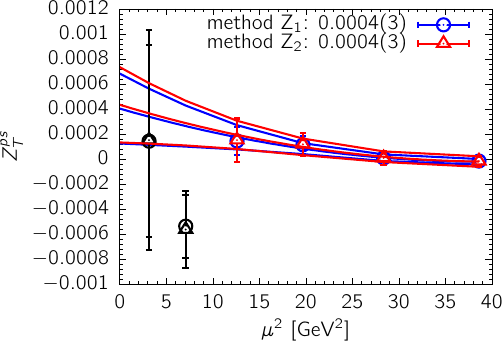}
  \includegraphics[width=0.24\textwidth]{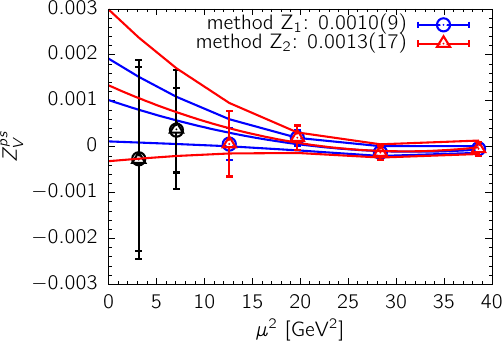}

  \includegraphics[width=0.24\textwidth]{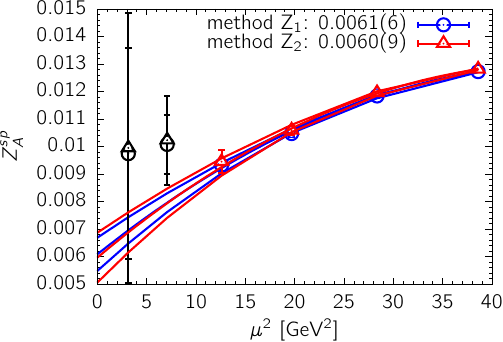}
  \includegraphics[width=0.24\textwidth]{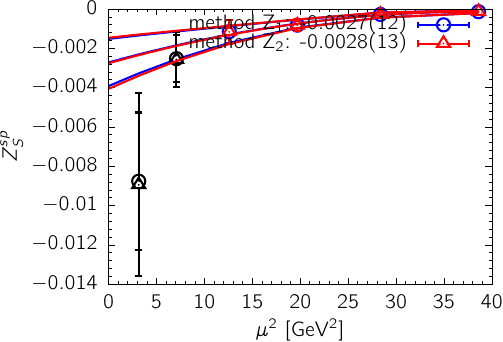}
  \includegraphics[width=0.24\textwidth]{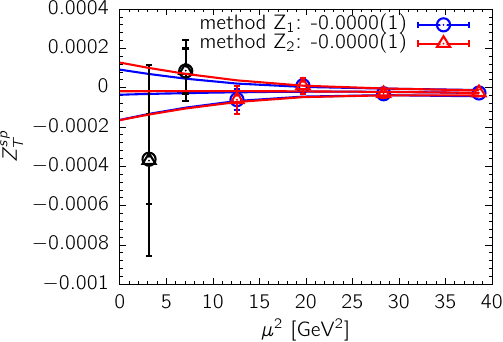}
  \includegraphics[width=0.24\textwidth]{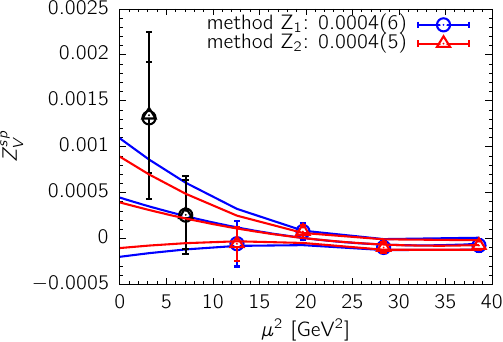}
  \caption{$Z_\Gamma$ in $\MSb$ at $2\GeV$ and quadratic
    extrapolation in $\mu^2$, calculated on a06m310 ensemble. The results from two
    different strategies, $Z_1$ (blue) and $Z_2$ (red) are being
    compared. Each column represents axial, scalar, tensor and vector
    operator, respectively. Each row represents one of the 5 nonzero
    matrix elements of $Z$ in $\{u-d,u+d,s\}$ in Eq.~\eqref{eq:Z_RI}.
  }  
  \label{fig:Z_MSb2GeV_a06}
\end{figure*}

The data for $Z_\Gamma$ in $\MSb$ at $2\GeV$ is shown in
Fig.~\ref{fig:Z_MSb2GeV_a06} for the $a06m310$ ensemble as a functions
of $\mu$ along with a quadratic extrapolation in $\mu^2$.  The results
for the diagonal parts of $Z_1$ and $Z_2 $ show a difference which
vanishes in the continuum limit.  The off-diagonal mixing elements
$Z_\Gamma^{u+d,s}$ and $Z_\Gamma^{s,u+d}$ shown in the bottom two rows
are all smaller than $1\%$ and the $Z_1$ and $Z_2$ results essentially
overlap. \looseness-1

\section{Chiral-Continuum-Finite-Volume (CCFV) extrapolation and results}
\label{sec:CCFV}

We have carried out four analyses. (i) Standard fits to remove ESC with $Z_1$ renormalization. (ii) Fits including $N\pi$ state to remove ESC with $Z_1$ renormalization.
(iii) Standard fits to remove ESC with $Z_2$ renormalization. (iv) Fits including $N\pi$ state to remove ESC with $Z_2$ renormalization. For details
on ESC fits see  Refs.~\cite{Gupta:2021ahb, Park:2023tsj}.

The renormalized axial, $g_A^{u,d,s}$, tensor, $g_T^{u,d,s}$, and
strange scalar, $g_S^s$ charges are extrapolated to the physical
point, $a \to 0$, $M_\pi = 135$~MeV, and $M_\pi L \to \infty$ using
the CCFV ansatz, $ g(a,M_\pi,M_\pi L) = c_0+c_a a+ c_2 M_\pi^2 + c_3
\frac{M_\pi^2 e^{-M_\pi L}}{\sqrt{M_\pi L}}$ that includes the leading
corrections in all three variables $\{a,M_\pi,M_\pi L\}$.  For the
scalar charges $g_S^{u,d}$, the leading pion mass dependence starts
$O(M_\pi)$ due to the pion loop and a different finite volume
correction~\cite{Gupta:2021ahb}, $g_S^{u,d}(a,M_\pi,M_\pi L) =d_0+d_a
a+ d_2 M_\pi+ d_3 M_\pi^2 + d_4 M_\pi \left(1-\frac{2}{M_\pi L}\right)
e^{-M_\pi L}$. We ignore higher order terms (chiral logs, etc.)
since we only have data at three values of $M_\pi$. Results (to be considered preliminary until published) for all 
four analyses are summarized in Table~\ref{tab:gAST-Z}. 

\begin{figure*}
  \center
  \includegraphics[width=0.23\textwidth]{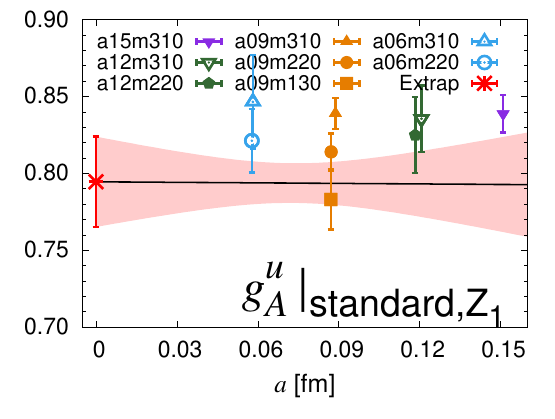}
  \includegraphics[width=0.23\textwidth]{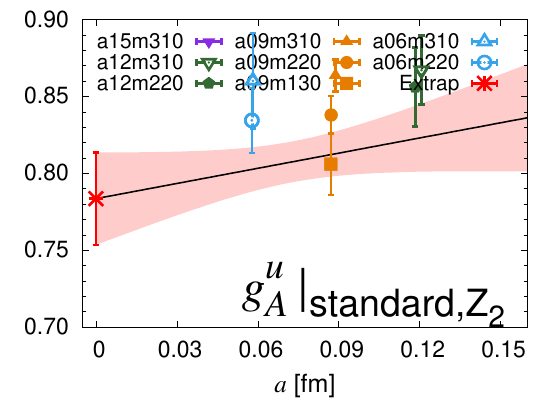}
  \includegraphics[width=0.23\textwidth]{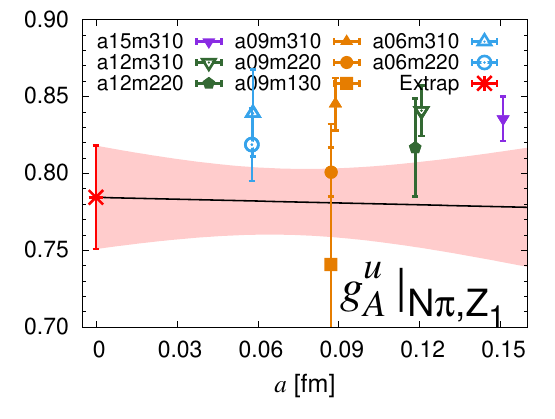}
  \includegraphics[width=0.23\textwidth]{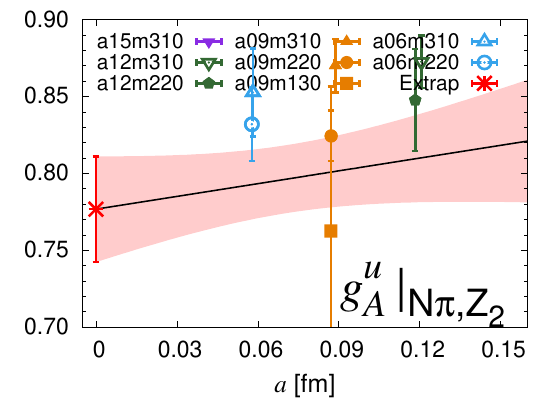}
  
  \includegraphics[width=0.23\textwidth]{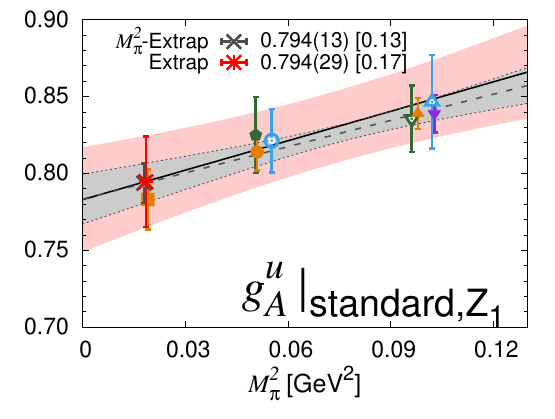}
  \includegraphics[width=0.23\textwidth]{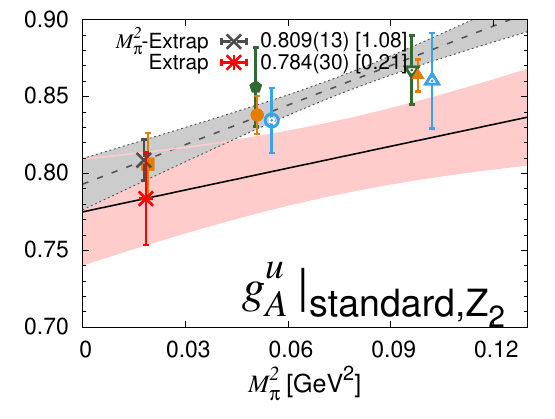}
  \includegraphics[width=0.23\textwidth]{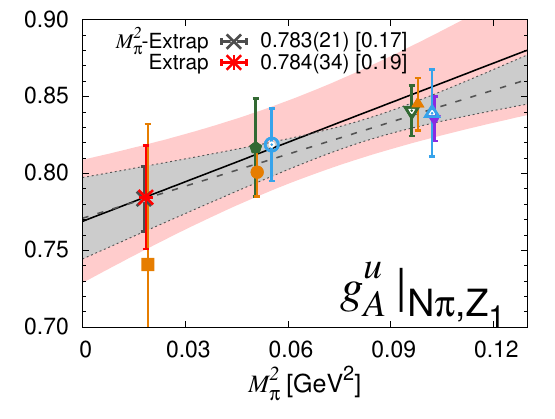}
  \includegraphics[width=0.23\textwidth]{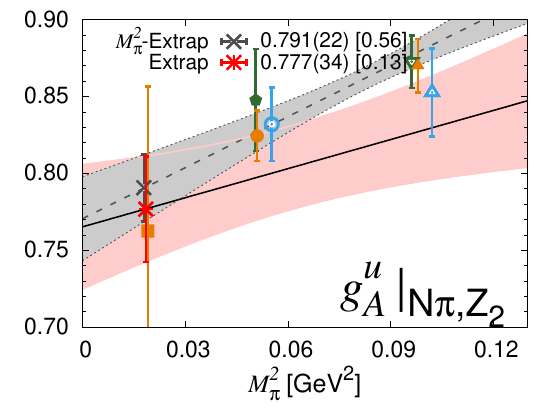}
  
  \includegraphics[width=0.23\textwidth]{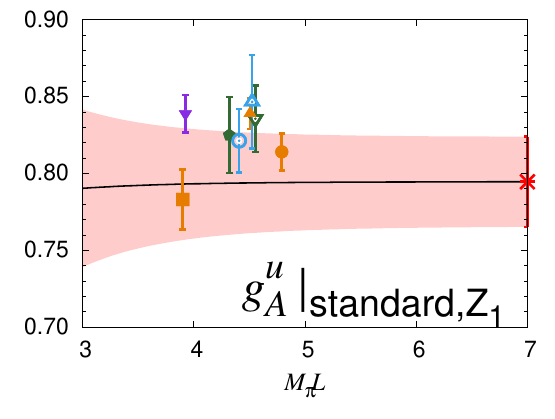}
  \includegraphics[width=0.23\textwidth]{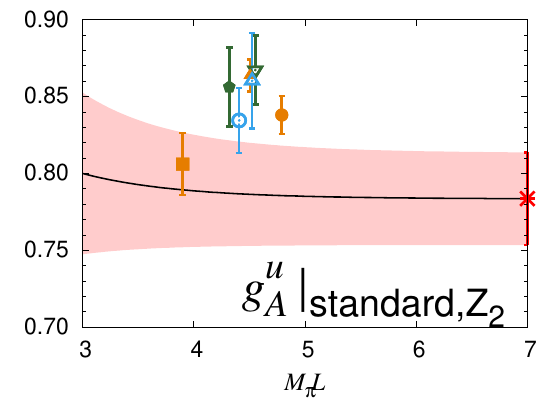}
  \includegraphics[width=0.23\textwidth]{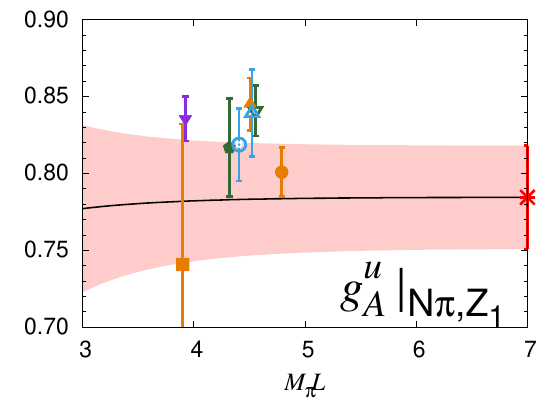}
  \includegraphics[width=0.23\textwidth]{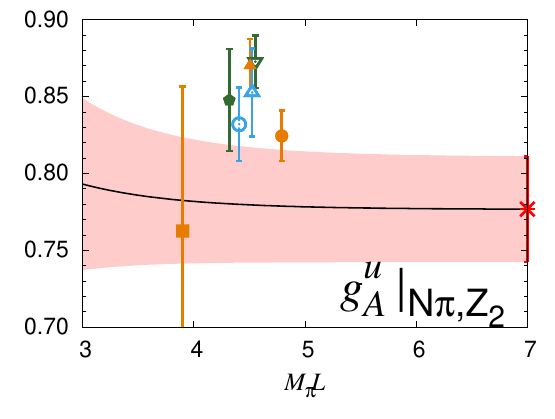}
   \caption{Each column shows the CCFV extrapolation for
    $g_A^{u}$ versus $a$, $M_\pi^2$ and $M_\pi L$ with the other 2 variables set to their physical point values. The rows give the 4 different analyses done: (i) \{standard, $Z_1$\},
    (ii) \{standard,$Z_2$\}, (iii) \{$N\pi$,$Z_1$\} and
    (iv) \{$N\pi$,$Z_2$\}. }
  \label{fig:CCFV_gAu} 
\end{figure*}

Figure \ref{fig:CCFV_gAu} shows CCFV fits to $g_A^u$. The data 
from two renormalization methods, Z1 and Z2 defined in
Sec.~\ref{sec:Z1Z2}, show different $a$ and $M_\pi$ dependence, but
the extrapolated results are consistent. This holds for all 
charges $g_{A,S,T}^{u,d,s}$. Similarly, results from the two excited state fit methods, `standard' and $N\pi$ are consistent for $g_{A,T}^{u,d,s}$, however, there 
is a significant difference between the `standard' and $N\pi$ results for the 
scalar charges $g_S^{u,d,s}$. Note that the ESC fits to 
the current data do not distinguish between the two on the basis of $\chi^2/dof$ but give different estimates. The large enhancement in
$g_S^{u,d}$ due to $N\pi$ state contribution is discussed in
Ref.~\cite{Gupta:2021ahb}.   For the strange charges $g_\Gamma^s$, we do
not expect a large contribution from multihadron states since the lowest
state is $\Sigma K$, which has a large mass gap ($> N(1440)$), so we consider the 
`standard' analysis more appropriate than `$N\pi$'.   

\begin{table*}
\centering
  \begin{tabular}{l|lll|lll}
    & \multicolumn{3}{c|}{Standard Method for removing ESC} & \multicolumn{2}{c}{$N \pi$ method}\\
    $q$ & $g_A^q$ & $g_T^q$ & $g_S^q$ & $g_A^q$  & $g_T^q$  &  $g_S^q$  \\
    \hline
    u & 0.794(29)    & 0.789(27)    & 6.48(66)     & 0.784(34)    & 0.788(37)    & 8.8(1.3)    \\
    d & -0.385(26)   & -0.203(11)   & 6.09(73)     & -0.416(36)   & -0.188(17)   & 8.69(89)    \\
    s & -0.051(11)   & -0.0016(11)  & 0.38(12)     & -0.066(12)   & -0.0016(11)  & 0.67(16)    \\
    \hline
    u & 0.784(30)    & 0.778(28)    & 6.45(68)     & 0.777(34)    & 0.780(37)    & 8.9(1.4)    \\
    d & -0.381(26)   & -0.201(12)   & 6.09(75)     & -0.414(37)   & -0.185(17)   & 8.75(91)    \\
    s & -0.053(11)   & -0.0015(12)  & 0.36(13)     & -0.069(13)   & -0.0015(12)  & 0.67(17)    \\
  \end{tabular}
  \caption{Preliminary results for flavor diagonal charges of proton (for neutron interchange $u \leftrightarrow d$) 
  with the two
    strategies used to remove ESC, "standard" and $N\pi$. The top three
    rows give the results obtained using the $Z_1$ renormalization
    method, and the bottom three with the $Z_2$ renormalization
    method.\looseness-1}
  \label{tab:gAST-Z}
\end{table*}

\section{Conclusions}
We have now determined all the nucleon flavor diagonal charges,
$g_{A,S,T}^{u,d,s}$, removing previous approximations in
renormalization and ESC fits made in Ref.~\cite{Lin:2018obj,Gupta:2018lvp}, which are shown to be negligible. Further details will be
provided in a paper under preparation.  A key issue with the ES analysis is the need for a data driven method  to distinguish
between the `standard' and `$N\pi$' fits, 
particularly for $g_{A,T}^{u,d}$. To address this limitation, we are increasing the statistics on two physical $M_\pi$ ensembles.  


\begin{acknowledgments}
We thank the MILC collaboration for providing the 2+1+1-flavor HISQ
lattices. The calculations used the Chroma software
suite~\cite{Edwards:2004sx}.
This research used resources at (i) the NERSC, a DOE Office of Science
facility supported under Contract No.\ DE-AC02-05CH11231; (ii) the
OLCF, a DOE Office of Science User Facility supported under Contract
DE-AC05-00OR22725, through ALCC awards LGT107 and INCITE awards PHY138
and HEP133; (iii) the USQCD collaboration, which is funded by the
Office of Science of the U.S. DOE; and (iv) Institutional Computing at
Los Alamos National Laboratory.
S.P. acknowledges support from the U.S. DOE Contract
No.\ DE-AC05-06OR23177, under which Jefferson Science Associates, LLC,
manages and operates Jefferson Lab. Also acknowledged is support from
the Exascale Computing Project (17-SC-20-SC), a collaborative effort
of the U.S. DOE Office of Science and the National Nuclear Security
Administration.
S.P. acknowledges the support of the DOE under contract
No.\ DE-AC52-07NA27344 (LLNL) with support from the ASC COSMON project.
T.B. and R.G. were partly supported by the U.S.\ DOE, Office of
Science, Office of High Energy Physics under Contract
No.\ DE-AC52-06NA25396. S.P., T.B., R.G., S.M. and B.Y. were partly
supported by the LANL LDRD program, and S.P. by the Center for
Nonlinear Studies.
\end{acknowledgments}

\bibliographystyle{JHEP}
\bibliography{ref} 

\providecommand{\href}[2]{#2}\begingroup\raggedright\begin{thebibliography}{10}

\bibitem{Bazavov:2012xda}
{\bf MILC} Collaboration, A.~Bazavov, C.~Bernard, J.~Komijani, C.~De{T}ar,
  L.~Levkova, W.~Freeman, S.~Gottlieb, R.~Zhou, U.~M. Heller, J.~E. Hetrick,
  J.~Laiho, J.~Osborn, R.~L. Sugar, D.~Toussaint, and R.~S. Van~de Water {\em
  Phys. Rev. D} {\bf 87} (2013), no.~5 054505,
  [\href{http://xxx.lanl.gov/abs/1212.4768}{{\tt 1212.4768}}].

\bibitem{Lin:2018obj}
H.-W. Lin, R.~Gupta, B.~Yoon, Y.-C. Jang, and T.~Bhattacharya {\em Phys. Rev.}
  {\bf D98} (2018), no.~9 094512,
  [\href{http://xxx.lanl.gov/abs/1806.10604}{{\tt 1806.10604}}].

\bibitem{Gupta:2018lvp}
R.~Gupta, B.~Yoon, T.~Bhattacharya, V.~Cirigliano, Y.-C. Jang, and H.-W. Lin
  {\em Phys. Rev.} {\bf D98} (2018), no.~9 091501,
  [\href{http://xxx.lanl.gov/abs/1808.07597}{{\tt 1808.07597}}].

\bibitem{Gupta:2021ahb}
R.~Gupta, S.~Park, M.~Hoferichter, E.~Mereghetti, B.~Yoon, and T.~Bhattacharya
  {\em Phys. Rev. Lett.} {\bf 127} (2021), no.~24 242002,
  [\href{http://xxx.lanl.gov/abs/2105.12095}{{\tt 2105.12095}}].

\bibitem{FLAG:2021npn}
{\bf Flavour Lattice Averaging Group (FLAG)} Collaboration, Y.~Aoki {\em
  et~al.} {\em Eur. Phys. J. C} {\bf 82} (2022), no.~10 869,
  [\href{http://xxx.lanl.gov/abs/2111.09849}{{\tt 2111.09849}}].

\bibitem{Park:2023tsj}
S.~Park, T.~Bhattacharya, R.~Gupta, H.-W. Lin, S.~Mondal, and B.~Yoon {\em PoS}
  {\bf LATTICE2022} (2023) 118, [\href{http://xxx.lanl.gov/abs/2301.07890}{{\tt
  2301.07890}}].

\bibitem{Martinelli:1994ty}
G.~Martinelli, C.~Pittori, C.~T. Sachrajda, M.~Testa, and A.~Vladikas {\em
  Nucl.Phys.} {\bf B445} (1995) 81--108,
  [\href{http://xxx.lanl.gov/abs/hep-lat/9411010}{{\tt hep-lat/9411010}}].

\bibitem{Sturm:2009kb}
C.~Sturm, Y.~Aoki, N.~H. Christ, T.~Izubuchi, C.~T.~C. Sachrajda, and A.~Soni
  {\em Phys. Rev.} {\bf D80} (2009) 014501,
  [\href{http://xxx.lanl.gov/abs/0901.2599}{{\tt 0901.2599}}].

\bibitem{Gracey:2011fb}
J.~A. Gracey {\em Eur. Phys. J.} {\bf C71} (2011) 1567,
  [\href{http://xxx.lanl.gov/abs/1101.5266}{{\tt 1101.5266}}].

\bibitem{Chetyrkin:1999pq}
K.~G. Chetyrkin and A.~Retey {\em Nucl. Phys.} {\bf B583} (2000) 3--34,
  [\href{http://xxx.lanl.gov/abs/hep-ph/9910332}{{\tt hep-ph/9910332}}].

\bibitem{Larin:1993tq}
S.~A. Larin {\em Phys. Lett. B} {\bf 303} (1993) 113--118,
  [\href{http://xxx.lanl.gov/abs/hep-ph/9302240}{{\tt hep-ph/9302240}}].

\bibitem{Green:2017keo}
J.~Green, N.~Hasan, S.~Meinel, M.~Engelhardt, S.~Krieg, J.~Laeuchli, J.~Negele,
  K.~Orginos, A.~Pochinsky, and S.~Syritsyn {\em Phys. Rev.} {\bf D95} (2017),
  no.~11 114502, [\href{http://xxx.lanl.gov/abs/1703.06703}{{\tt 1703.06703}}].

\bibitem{Gracey:2020rok}
J.~A. Gracey {\em Phys. Rev. D} {\bf 102} (2020), no.~3 036002,
  [\href{http://xxx.lanl.gov/abs/2001.11282}{{\tt 2001.11282}}].

\bibitem{Gracey:2000am}
J.~A. Gracey {\em Phys.Lett.} {\bf B488} (2000) 175--181,
  [\href{http://xxx.lanl.gov/abs/hep-ph/0007171}{{\tt hep-ph/0007171}}].

\bibitem{Bhattacharya:2016zcn}
T.~Bhattacharya, V.~Cirigliano, S.~D. Cohen, R.~Gupta, H.-W. Lin, and B.~Yoon
  {\em Phys. Rev.} {\bf D94} (2016), no.~5 054508,
  [\href{http://xxx.lanl.gov/abs/1606.07049}{{\tt 1606.07049}}].

\bibitem{Yoon:2016jzj}
B.~Yoon {\em et~al.} {\em Phys. Rev. D} {\bf 95} (2017), no.~7 074508,
  [\href{http://xxx.lanl.gov/abs/1611.07452}{{\tt 1611.07452}}].

\bibitem{Park:2021ypf}
{\bf Nucleon Matrix Elements (NME)} Collaboration, S.~Park, R.~Gupta, B.~Yoon,
  S.~Mondal, T.~Bhattacharya, Y.-C. Jang, B.~Jo\'o, and F.~Winter {\em Phys.
  Rev. D} {\bf 105} (2022), no.~5 054505,
  [\href{http://xxx.lanl.gov/abs/2103.05599}{{\tt 2103.05599}}].

\bibitem{Edwards:2004sx}
{\bf SciDAC, LHPC, UKQCD} Collaboration, R.~G. Edwards and B.~Jo{\'o} {\em
  Nucl. Phys. Proc. Suppl.} {\bf 140} (2005) 832,
  [\href{http://xxx.lanl.gov/abs/hep-lat/0409003}{{\tt hep-lat/0409003}}].

\end{thebibliography}\endgroup

\end{document}